\documentstyle[aps,prd,12pt,tighten,amssymb,eqsecnum,axodraw,cite]{revtex}

\title{\vspace*{1.5in}Gauge Coupling Instability and Dynamical Mass
  Generation in \({\mathcal N}\!=\!1\) Supersymmetric QED$_3$}

\author{A. Campbell--Smith${}^{a,b}$,
  N.E. Mavromatos${}^{a,b}$ and J. Papavassiliou${}^{a}$}

\address{${}^a$ Theory Division, CERN, CH--1211 Geneva 23,
  Switzerland.\\${}^b$Theoretical Physics (University of
  Oxford), 1 Keble Road, OXFORD, OX1 3NP, U.K.}

\newcommand{\bard}{d \!\!\rule[6.5pt]{4pt}{0.3pt}}

\begin{document}

\maketitle


\vspace*{1cm}
\begin{abstract}
  Using superfield Dyson--Schwinger equations, we compute the infrared
  dynamics of the semi--amputated full vertex, corresponding to the
  effective running gauge coupling, in \(N\)--flavour \({\mathcal
    N}\!=\!1\) supersymmetric QED$_3$.  It is shown that the presence
  of a supersymmetry--preserving mass for the matter multiplet
  stabilizes the infrared gauge coupling against oscillations present
  in the massless case, and we therefore infer that the massive vacuum
  is thus selected at the level of the (quantum) effective action.  We
  further demonstrate that such a mass can indeed be generated
  dynamically in a self--consistent way by appealing to the superfield
  Dyson--Schwinger gap equation for the full matter propagator.
\end{abstract}

\vspace*{-5in}
\begin{flushright}
CERN--TH/99--141\\
OUTP--99--24P\\
hep--th/9905132
\end{flushright}
\vspace*{4.5in}


\section{Introduction.}
\label{sec:intro}

Three dimensional U(1) gauge field theories, whether as toy models of
four dimensional physics or as interesting physical theories in their
own right, have attracted considerable attention
\cite{pisarski84,appelquist86,pennington91,kondo+nak92,dorey92,maris96,aitch+mav:prb,aitchetal97}.
In particular, they have been shown to exhibit a rich and non--trivial
infrared structure, exhibiting dynamical mass generation and critical
behaviour; the study of these phenomena requires non--perturbative
techniques.  The question of how the incorporation of supersymmetry
affects the infrared physics of these theories has also been addressed
\cite{pisarski84,koopmans89,adrian99a} and it has been found that when
\({\mathcal N}\!=\!1\) mass can be generated dynamically without
breaking the supersymmetry, while mass generation in the \({\mathcal
  N}\!=\!2\) model is forbidden.  For the \({\mathcal N}\!=\!1\) model
an important question remains unanswered: whether the massive or
massless solutions are selected by the theory.  In non--supersymmetric
theories this question can be addressed by appealing to the effective
potential and energetic favourability \cite{pisarski84,appelquist86};
in supersymmetric theories the effective potential vanishes and this
argument fails.  It has been suggested \cite{pisarski84} that the
issue can be resolved in supersymmetric theories at the level of the
(quantum) effective action (which does not vanish) and the Ward
identities arising from supersymmetry.  The Ward identities have been
found to play a crucial r\^{o}le in this selection in models which
have additional constraints \cite{alvarez78,diamandis98}, but do not
yield any extra information for the unconstrained U(1) gauge field
theory \cite{adrian99a}.  In this paper we propose a resolution of
this problem for the basic $N$--flavour \({\mathcal N}\!=\!1\) U(1)
theory at the level of the effective action: by considering an
appropriate set of Dyson--Schwinger equations we show that a mass can
be self--consistently generated without breaking supersymmetry, and
that this mass stabilizes the infrared effective gauge coupling which
in the massless case oscillates without definite sign.

We adopt a superfield formalism throughout this paper, and work with a
large number $N$ of matter flavours.  We consider the superfield
Dyson--Schwinger equation for the semi--amputated full three--point
vertex \cite{papavassiliou91,cornwall95,mavro99} which corresponds to
the effective (running) gauge coupling \cite{cornwall89}, and compute
the infrared dynamics of this vertex in the presence of vanishing and
non--vanishing masses for the matter multiplet.  In the massless case
the dynamics are described by a non--linear differential equation of
Emden--Fowler type \cite{kamke59,emden:fowler}, which admits only
oscillatory solutions in the infrared; we interpret this as indicating
instabilities in the (quantum) effective action.  Repeating the
computation with the insertion (by hand) of a finite mass for the
matter multiplet we find that these instabilities disappear, and the
solution shows the existence of a non--trivial infrared fixed point
for the running coupling.  We then demonstrate that a mass for the
matter superfield can be dynamically generated in a self--consistent
way in this approach by appealing to the Dyson--Schwinger equation for
the full matter propagator cast in terms of the semi--amputated full
vertex.  In agreement with reference \cite{adrian99a} we find no
evidence for a critical flavour number, above which dynamical mass
generation does not occur.

In reference \cite{adrian99a} the gap equation for the full matter
propagator was studied incorporating a full vertex which by
construction satisfied the U(1) Ward identity; instead, in this work,
the form of the full vertex in the deep infrared will be determined
self--consistently from the Dyson--Schwinger equation.  The truncated
form of the Dyson--Schwinger equation is not manifestly gauge
invariant, {i.e.} it does not respect the Ward identity for general
incoming momenta and there is a residual explicit dependence on the
covariant gauge fixing parameter.  However in the physical, on--shell
limit of vanishing incoming momenta we find that the Ward identity is
satisfied and that the dependence on the gauge fixing parameter drops
out.  Since in the present work we are interested only in obtaining
and analysing the structure of the vertex in the deep infrared, the
lack of gauge invariance away from vanishing incoming momentum will
not adversely affect our conclusions.  The superfield formalism we
adopt keeps supersymmetry manifest and thereby radically simplifies
the system of equations one would obtain for the full vertices in a
component computation.  The disadvantages of the superfield formalism
\cite{clark77,clark+love} lie in the propagation of gauge artifacts in
the connexion superfield which result in spurious infrared
divergences.  By working in a general (covariant) gauge we will have
enough flexibility to remove these divergences by an appropriate gauge
choice.

This paper is organized as follows: in section \ref{sec:action} we
construct the action functional for \mbox{\({\mathcal N}\!=\!1\)}
supersymmetric QED$_3$ in superfield formalism and give the form of
the dressed propagators for the matter and connexion superfields.  In
section \ref{sec:dsv} we introduce the superfield semi--amputated
vertex and construct its Dyson--Schwinger equation; we then solve the
Dyson--Schwinger equation in the presence of vanishing and
non--vanishing masses for the matter superfield and interpret the
results.  In section \ref{sec:z} we discuss the infrared properties
of the matter and connexion superfield propagators, in association
with approximations made in section \ref{sec:dsv}; we show in section
\ref{sec:mass} that when the previously computed vertex is
incorporated in the gap equation for the matter propagator, a mass is
dynamically generated self--consistently.  Finally we present our
conclusions in section \ref{sec:conc}.

\section{The Action.}
\label{sec:action}

We consider a model with ${\mathcal N}=1$ supersymmetry, $N$ matter
flavours and local U(1) gauge invariance.  The required action
functional then comprises three parts: the gauge invariant classical
field strength term for the (spinor) connexion $\Gamma_\alpha$, a
(Lorentz) gauge fixing term, and a locally U(1) invariant kinetic term
for the matter superfields $\Phi$ and $\Phi^*$:
\begin{equation} \label{n=1action}
S = S_{g}^{\mathrm{class}} + S_{g}^{\mathrm{GF}} + S_m ;
\end{equation}
\begin{eqnarray}
S_g^{\mathrm{class}} &=& \int d^3 x\, d^2\theta \; \Gamma_\alpha \left(
-\frac{1}{8} D^\eta D^\alpha D^\beta D_\eta \right) \Gamma_\beta
,\nonumber\\
S_g^{\mathrm{GF}} &=& \int d^3 x\, d^2 \theta\; \Gamma_\alpha \left(
\frac{1}{4\xi} D^\alpha D^2 D^\beta \right) \Gamma_\beta ,\nonumber\\
S_m &=& \int d^3 x\, d^2\theta\; \left(-\frac{1}{2}\right) \left[
\nabla^\alpha \Phi \right]^* \left[ \nabla_\alpha \Phi \right] .
\end{eqnarray}
We have included in the matter part an implicit sum over $N$ flavours,
which do not interact with each other directly, but interact with the
same connexion superfield.  The U(1) covariant derivative
$\nabla^\alpha$ is given by \[ \nabla^\alpha \doteq D^\alpha -ie
\Gamma^\alpha,\] where $e$ is the (dimensionful) gauge coupling.  We
work within large--$N$, in which the quantity \[\frac{e^2 N}{4} =
\alpha \] is kept fixed (but large) as $N$ becomes large.  From the
action (\ref{n=1action}) it is easy to derive the dressed propagators
for the matter and connexion superfields:
\begin{eqnarray} \label{props}
\Delta(p;12) &=& \frac{i}{A(p)} \frac{D^2 (p) -
M(p)}{p^2+M^2(p)} \delta^2(12), \nonumber\\
\Delta_{\alpha\beta} (p;12) &=& -i
\frac{1}{p^4}\frac{1}{B(p)}\left[ (1\!+\!\xi) \, p_{\alpha\beta}
\, D^2 - (1\!-\!\xi) \, C_{\alpha\beta} \, p^2 \right] \delta^2(12).
\end{eqnarray}
The matter superfield contains a possible dynamically generated mass
function \(M(p)\) and the scalar functions \(A\) and \(B\)
parameterize the dressing of the matter and connexion superfields
respectively.  The unknown functions $A$, $B$, and $M$ can at least in
principle be determined from the appropriate Dyson--Schwinger
equations; we return to these in sections \ref{sec:z} and
\ref{sec:mass}.  The dressed three--point vertex derived from the
action above is shown in figure \ref{fig:frv}.
\begin{figure}[thb]
\begin{center}
\begin{picture}(200,80)
\Line(0,30)(60,30)
\Vertex(30,30){4}
\Photon(30,34)(50,70){3}{3.5}
\LongArrow(5,25)(20,25)
\Text(15,17)[r]{$p$}
\LongArrow(40,25)(54,25)
\Text(45,17)[c]{$q$}
\LongArrow(28,45.445)(35.28,58.555)
\Text(27,56)[r]{$p\!-\!q$}
\Text(0,37)[l]{$\Phi$}
\Text(59,37)[]{$\Phi^*$}
\Text(55,65)[l]{$\Gamma_\alpha$}
\Text(80,30)[l]{$: \; -\frac{e}{2} \,G (p,p\!-\!q,q) \,C^{\alpha\beta}
  \,D_\beta (q)$}
\end{picture}
\caption{Structure and momentum assignments of the dressed
  three--point vertex; solid lines represent matter superfields and
  wavy lines the connexion superfield.\label{fig:frv}}
\end{center}
\end{figure}

In this paper we use a convenient spinor index notation for
three--vectors, in which they are represented as symmetric second rank
spinors; spinor indices are raised and lowered by the antisymmetric
metric $C$.  We collect here some useful identities which will be used
in what follows \cite{gates83:super} (\(\{\partial_\mu , \theta^\nu \}
= \delta_\mu^\nu \)):
\begin{eqnarray}
  \label{suspids}
  && A^\alpha = C^{\alpha\beta} A_\beta,\nonumber\\
  && A_\beta = A^\alpha C_{\alpha\beta} \nonumber\\
  && A^2 \doteq \frac{1}{2} A^\alpha A_\alpha,\nonumber\\ 
  && C_{\mu\nu} C^{\alpha\beta} = \delta_{[\mu}^{\alpha}\,
  \delta_{\nu]}^{\beta},\nonumber\\
  && p_{\mu\nu} q^{\nu\alpha} = \delta_\mu^\alpha \, p\cdot q,\nonumber\\ 
  && D^\mu (q) = \partial^\mu + \theta^\nu q_{\mu\nu},\nonumber\\ 
  && D^2 (q) = \partial^2 + \theta^\mu q_{\mu\nu} \partial^\nu + q^2
  \theta^2,\nonumber\\ 
  && D^\mu (q) D^2(q) = q^{\mu\lambda} D_\lambda (q),\nonumber\\ 
  && D^\mu (q) D^\nu (q) = q^{\mu\nu} + C^{\nu\mu} D^2(q).
\end{eqnarray}

In the next section we construct and analyse the truncated
Dyson--Schwinger equation for the full three--point vertex shown in
figure \ref{fig:frv} with and without a mass for the matter multiplet
and demonstrate that oscillations present in the massless case
disappear when a mass for the matter multiplet is included.

\section{Dyson--Schwinger Equation for the Vertex.}
\label{sec:dsv}

Throughout this paper we use a truncation to leading order in $1/N$;
the truncated Dyson--Schwinger equation for the full vertex is shown
schematically in figure \ref{fig:dsv}.

\begin{figure}[thb]
\begin{center}
\begin{picture}(300,65)(0,0)
\Line(0,10)(31,29)
\GCirc(34,30){4}{0}
\Line(37,29)(68,10)
\Photon(34,64)(34,34){3}{3}
\Text(83,30)[]{$=$}
\Line(98,10)(128,30)
\GCirc(129,30){1}{0}
\Line(130,30)(160,10)
\Photon(129,64)(129,31){3}{3}
\Text(175,30)[]{$+$}
\Line(190,10)(221,29)
\Line(227,29)(258,10)
\GCirc(224,30){4}{0}
\Photon(224,64)(224,34){3}{3}
\PhotonArc(224,50)(40,238.33,301.67){3}{3.5}
\GCirc(200,16.451){4}{0}
\GCirc(248,16.451){4}{0}
\GCirc(212,23.23){4}{0.5}
\GCirc(236,23.23){4}{0.5}
\GCirc(224,10){4}{0.5}
\end{picture}
\caption{Schematic form of the Dyson--Schwinger  equation for the
  full vertex.  Solid lines represent matter superfields, and wavy
  lines the connexion superfield.  Blobs indicate full non--perturbative quantities.}
\label{fig:dsv}
\end{center}
\end{figure}

The graph on the left hand side of figure \ref{fig:dsv} is written as follows
\begin{equation}
  \label{graphlhs}
  \left( -\frac{e}{2}G(p,p-q,q) \right) \int d^2 \theta\;
  \Phi(-p,\theta) \Gamma_\alpha (p-q,\theta)\,D^\alpha (q)
  \Phi^* (q,\theta),
\end{equation}
where we have assumed that we can perform the factorization into a
scalar function $G$ multiplying the superspace structure shown above;
the first graph on the right of figure \ref{fig:dsv} is of the same
form but with $G \mapsto 1$.  This factorization is similar in effect
to the approach taken in non--supersymmetric QED${}_3$ in which the
vertex is assumed to be a scalar function multiplying the usual Dirac
matrix.  This approximation is computationally convenient, but does
not satisfy the Ward identity except at vanishing incoming momentum,
where the full vertex becomes on--shell.  Given this limitation we
will only be interested in the behaviour of the theory in this regime;
as we will show, even with this restriction we can obtain interesting
information about the infrared fixed point structure.  As we will see
in section \ref{sec:mass}, the present approach gives rise to the same 
qualitative picture as that obtained in reference \cite{adrian99a},
namely that a mass can be generated dynamically for the matter
multiplet without breaking supersymmetry.

Following references \cite{papavassiliou91,cornwall95,mavro99} we
define the semi--amputated full non--perturbative vertex \(\hat{G}\)
as
\begin{equation}
  \label{ghatdef}
  \hat{G}(p_1,p_2,p_3) \doteq Z(p_1,p_2,p_3) \, G(p_1,p_2,p_3),
\end{equation}
where $Z$ is defined in terms of the functions $A$ and $B$ appearing
in the dressed propagators (\ref{props}) as
\begin{equation}
  \label{zdef}
  Z(p_1,p_2,p_3) \doteq A^{-1/2} (p_1) \, B^{-1/2} (p_2) \, A^{-1/2}
  (p_3) \geqslant 0.
\end{equation}
The quantity \(e \hat{G}\) is the appropriate and natural
generalization of the running charge in super--renormalizable gauge
field theories \cite{cornwall89}.  This definition is the same as the
generalization of the running charge in non--supersymmetric QED$_3$
\cite{mavro99}.  As we will show, this definition also simplifies the
structure of the integral equation for the vertex function
considerably.

\subsection{Vertex With Vanishing Mass.}
\label{sec:m=0v}

The one--loop graph on the right of figure \ref{fig:dsv} reads as follows:
\begin{eqnarray}
  \label{1loop}
  -\frac{e}{2}\left(-i\frac{e^2}{4}\right) \int &&\bard^3 k \,d^2\theta_1
   d^2\theta_2 d^2\theta_3\;G^3\;
   \left[ \frac{D^\mu (p+k) D^2 (p+k)}{A(p+k)\,(p+k)^2} \delta^2(12)
   \right] \times \nonumber\\
&& \times \left[ \frac{D^\alpha (q+k) D^2 (q+k)}{A(q+k)\,(q+k)^2} \delta^2(23)
   \right] \left[ \frac{(1\!+\!\xi) k_{\mu\nu} - (1\!-\!\xi)
   C_{\mu\nu} k^2}{B(k)\, k^4} \delta^2(31) \right]\times\nonumber\\
\rule{0pt}{14pt}&&\times \Phi(-p,\theta_1)\,\Gamma_\alpha (p-q,\theta_2)\, D^\nu (q)
   \Phi^* (q,\theta_3),
\end{eqnarray}
where
\begin{equation}
  \label{g3def}
G^3 = G(p,-k,p+k)\,G(p+k,p-q,q+k)\,G(q+k,k,q).  
\end{equation}
Following \cite{mavro99} we consider the external connexion superfield
momentum to be vanishingly small, leaving one external scale $p$ in
the problem; computing the superspace part of equation (\ref{1loop})
using the identities (\ref{suspids}) we obtain an integral equation
for the full vertex $G$.  Multiplying this equation through by $Z(p)$
we obtain the desired integral equation for the semi--amputated vertex
$\hat{G}$ from the Dyson--Schwinger equation in figure \ref{fig:dsv}:
\begin{eqnarray}
  \label{inteqm=0:a}
  \hat{G}(p) = Z(p) + \frac{1}{4} e^2 \int \bard^3 k \; \hat{G}^3 (k) \,
\frac{ (1\!+\!\xi) k\cdot(p+k) + (1\!-\!\xi) k^2}{(p+k)^2 \, k^4}.
\end{eqnarray}
We have made the approximation that $\hat{G}^3$ is the cube of one
scalar function dependent on the scale $k$ only \cite{mavro99}: this
is justified by the self--consistency of our results.  Note that in
the absence of the inhomogeneous term $Z$ on the right hand side of
equation (\ref{inteqm=0:a}) one obtains a integral equation involving
only $\hat{G}$, which can in principle be solved.  In the following we
will drop this inhomogeneous term, returning in section \ref{sec:z} to
a discussion of why this may be done with safety.

The angular integration is easy to perform, leaving the following
equation, which we have recast in the dimensionless variables $x \doteq
p/\alpha$, $y \doteq k/\alpha$:
\begin{eqnarray}
  \label{inteqm=0}
  \hat{G}(x) = \frac{1}{4\pi^2 N} \int dy\; \frac{\hat{G}^3}{y^2}
  \left\{ (1\!+\!\xi) \left( 1 + \frac{y^2-x^2}{2xy} \ln\left|
  \frac{y+x}{y-x} \right| \right) + (1\!-\!\xi) \frac{y}{x} \ln \left|
  \frac{y+x}{y-x} \right| \right\}.
\end{eqnarray}

Since we are interested in the deep infrared we consider the limit
$x\ll 1$, expand the logarithms in the above equation to second order
and obtain
\begin{equation}
  x^2 \hat{G} \simeq \frac{(2\!-\!\xi)}{3\pi^2 N} \int_0^x dy\;
  \hat{G}^3 (y) + \frac{x^2}{\pi^2 N} \int_x^\infty dy\; \frac{\hat{G}^3
  (y)}{y^2} + \frac{x^4 \xi}{3\pi^2 N} \int_x^\infty dy\; \frac{\hat{G}^3
  (y)}{y^4}.
\end{equation}
We drop the last term in the limit $x\ll 1$ and after appropriate
differentiations with respect to $x$ we arrive the equivalent
differential equation:
\begin{equation}
  \label{m=0de}
  x^3 \hat{G}'' + \left[ 3x^2 + \frac{x(1\!+\!\xi)}{\pi^2 N} \hat{G}^2 \right]
  \hat{G}' + \frac{5-\xi}{3\pi^2 N} \hat{G}^3 = 0.
\end{equation}
In the gauge \(\xi=-1\) and changing variables \(x\mapsto w= x^{-2}\)
the above equation can be recast as a differential equation of
Emden--Fowler type \cite{kamke59,emden:fowler}:
\begin{equation}
  \frac{d^2}{dw^2}{\hat{G}} + \frac{1}{2\pi^2 N} w^{-3/2} \hat{G}^3 = 0;
\end{equation}
in the limit $w\rightarrow \infty$ it has been shown
\cite{emden:fowler} that the only (real and non--divergent) solutions
of this equation are oscillatory.  We interpret this as indicating
that in the absence of a mass for the matter multiplet the gauge
coupling is subject to instabilities which render it unphysical.  In
the next subsection we study the effects of a mass introduced by hand
and show that these instabilities are removed, and that the coupling
is driven to an non--trivial infrared fixed point.

\subsection{Vertex With Non--Vanishing Mass.}
\label{sec:m=mv}

In this subsection we repeat the computation above retaining the
effects of a constant mass $M(p) \simeq M(0) = M\neq 0$ in the
propagators (\ref{props}).  Here we will put in the mass by hand, in
section \ref{sec:mass} we will show that the semi--amputated full
vertex we consider admits the dynamical generation of a mass.  The
one--loop graph on the right of figure \ref{fig:dsv} now reads
\begin{eqnarray}
  \label{1loopM}
  -\frac{e}{2}\left(-i\frac{e^2}{4}\right) \int &&\bard^3 k\,d^2\theta_1
   d^2\theta_2 d^2\theta_3\;  G^3\;
   \left[ \frac{D^\mu (p+k) \left(D^2
   (p+k)-M\right)}{A(p+k)\,\left((p+k)^2+M^2\right)} \delta^2(12)
   \right] \times\nonumber\\
&&\times \left[ \frac{D^\alpha (q+k) \left(D^2 (q+k) -
   M\right)}{A(q+k)\,\left((q+k)^2 + M^2\right)} \delta^2(23)
   \right] \left[ \frac{(1\!+\!\xi) k_{\mu\nu} - (1\!-\!\xi)
   C_{\mu\nu} k^2}{B(k)\, k^4} \delta^2(31) \right]\times \nonumber\\
\rule{0pt}{14pt}&&\times\Phi(-p,\theta_1)\,\Gamma_\alpha (p-q,\theta_2)\, D^\nu (q) \Phi^* (q,\theta_3),
\end{eqnarray}
where as before $G^3$ is given by equation (\ref{g3def}).  Again the
superspace parts of this equation can be evaluated using the
identities (\ref{suspids}); in contrast to the case of
non--supersymmetric QED${}_3$, where the inclusion of a mass for the
fermions significantly alters the structure of the integral equation,
here the mass terms only appear in the denominators of the matter
propagators, for the terms linear in $M$ in the numerator cancel.  In
this respect, a mass for the matter superfield behaves like a trivial
infrared regulator, similar to the effect of a photon mass in the
non--supersymmetric model \cite{mavro99}.  On considering the external
connexion momentum vanishingly small and multiplying through by $Z(p)$
as before, the integral equation to be compared with
(\ref{inteqm=0:a}) reads
\begin{equation}
  \label{inteqm=ma}
  \hat{G}(p) = Z(p) + \frac{1}{4} e^2 \int \bard^3 k \; \hat{G}^3 (k) \,
\frac{ (1\!+\!\xi) k\cdot(p+k) + (1\!-\!\xi) k^2}{\left((p+k)^2 +M^2\right)\, k^4}.
\end{equation}
As before, it is easy to perform the angular integration, the result
of which reads (in dimensionless variables \(x\doteq p/\alpha\),
\(y\doteq k/\alpha\) and \(m\doteq M/\alpha\))
\begin{eqnarray}
  \label{inteqm=mb}
  \hat{G}(x) = \frac{1}{4\pi^2 N} \int dy\; \frac{\hat{G}^3}{y^2}
  &&\left\{(1\!+\!\xi) \left( 1 + \frac{y^2-x^2-m^2}{4xy} \ln\left[
  \frac{(y+x)^2 +m^2}{(y-x)^2+m^2} \right] \right) \right.\nonumber\\
&& \quad + \left.(1\!-\!\xi) \frac{y}{2x} \ln\left[
  \frac{(y+x)^2 +m^2}{(y-x)^2+m^2} \right] \right\}.
\end{eqnarray}
We have again dropped the inhomogeneous term $Z$ (see section
\ref{sec:z}); considering the deep infrared limit $x\ll 1$, and now
also $x\ll m$, we can expand the logarithms above to second order to
obtain the approximate form:
\begin{eqnarray}
  \label{inteqm=mc}
  \hat{G} \simeq \frac{(3\!-\!\xi)}{4\pi^2 N}\frac{1}{m^2} &&\int_0^x dy\;
  \hat{G}^3 (y) - \frac{(1\!+\!\xi)}{4\pi^2 N}\frac{x^2}{m^2} \int_0^x
  dy\; \frac{\hat{G}^3 (y)}{y^2} + \frac{1}{\pi^2 N} \int_x^\infty dy\;
  \frac{\hat{G}^3 (y)}{y^2 +m^2} \nonumber\\
- \frac{(1\!+\!\xi)}{4\pi^2 N} x^2 &&\int_x^\infty dy\; \frac{\hat{G}^3
  (y)}{y^2 \left( y^2 +m^2 \right)}.
\end{eqnarray}
Differentiation with respect to $x$ yields the integral--differential
equation
\begin{eqnarray}
  \hat{G}'(x) =&& \frac{(3\!-\!\xi)}{4\pi^2 N}\frac{\hat{G}^3(x)}{m^2}
- \frac{1}{\pi^2 N} \frac{\hat{G}^3(x)}{x^2+m^2}
- \frac{(1\!+\!\xi)}{4\pi^2 N} \frac{\hat{G}^3(x)}{m^2}
+ \frac{(1\!+\!\xi)}{4\pi^2 N} \frac{\hat{G}^3(x)}{x^2+m^2}\nonumber\\
&&- \frac{(1\!+\!\xi)}{2\pi^2 N} \frac{x}{m^2} \int_0^x dy\;
\frac{\hat{G}^3(y)}{y^2}
- \frac{(1\!+\!\xi)}{2\pi^2 N} x \int_x^\infty dy\;
\frac{\hat{G}^3(y)}{y^2 \left(y^2+x^2\right)} .
\end{eqnarray}
By the convenient choice of gauge $\xi=-1$ this can be reduced to a
first order non--linear differential equation, which can be integrated
with ease to yield
\begin{equation}
  \label{vsoln}
  \hat{G}(x) = \frac{1}{\left[ c + \frac{2}{\pi^2 N m^2} \left
  ( m \,\arctan\left(\frac{x}{m}\right) -x \right) \right]^{1/2}},
\end{equation}
where $c$ is an integration constant to be determined from the
boundary condition obtained from the original integral equation.  In
the limit $x\rightarrow 0$ the renormalization group $\beta$ function
vanishes:
\begin{equation}
  \lim_{x\rightarrow 0} x \frac{d\hat{G}(x)}{dx} = \lim_{x\rightarrow
  0} \frac{1}{\pi^2 N} \frac{\hat{G}^3(x) \, x^3}{m^2
  \left(x^2+m^2\right)} \rightarrow 0,
\end{equation}
and hence there is a non--trivial ($N$--independent) fixed point at
$x=0$ given by
\begin{equation}
  \hat{G}(0) = c^{-1/2}.
\end{equation}
Returning to the integral equation (\ref{inteqm=ma}) and taking the
limit $x\rightarrow 0$ we can investigate the constraints on the
integration constant $c$:
\begin{eqnarray}
  \label{bca}
  \hat{G}(0) = \frac{1}{\pi^2 N} \int_0^\infty dy\; \frac{\hat{G}^3(y)}{y^2+m^2}.
\end{eqnarray}
Since the right hand side of this equation is manifestly positive
definite, the trivial solution \(\hat{G}(0) = 1/\sqrt{c} =0\) is ruled
out, unless \(\hat{G}\) is trivially zero.  Noting that in the small
$x$ limit $\hat{G}$ differs from its fixed point value by a quantity
of order ${\mathcal O}(x^3)$, we can crudely approximate the integral
in (\ref{bca}) as follows
\begin{equation}
  \hat{G}(0) = \frac{1}{\pi^2 N} \int_0^m dy\;
  \frac{\hat{G}^3(0)}{y^2+m^2} + \frac{1}{\pi^2 N} \int_m^\infty dy\;
  \frac{1}{y^2+m^2},
\end{equation}
where we have set $\hat{G}$ in the second integral to its ultraviolet
asymptote of unity: both of these approximations are underestimates,
and therefore after performing the integrations we have
\begin{equation}
  \label{bcb}
  4\pi m \hat{G}(0) > \frac{1}{N} \left( 1 + \hat{G}^3 (0) \right).
\end{equation}
In principle for a given $m$ this leads to a (small) critical
coupling, below which there is no mass generation.  Note that even on
restoring the inhomogeneous term $Z$ (see section \ref{sec:z}) the
inequality is only modified to
\begin{equation}
  4\pi m \left(\hat{G}(0)-Z(0)\right)  > \frac{1}{N} \left( 1 +
  \hat{G}^3 (0) \right).
\end{equation}
Since at the level of the original integral equation (\ref{inteqm=ma})
in the gauge $\xi=-1$ it is obvious that $\hat{G} \geqslant Z$, we
obtain again a (small) critical coupling.  Note that in the above
analysis there is no evidence for a critical flavour number.

To conclude this section we should check our assertion that the value
of $\hat{G}$ at vanishing $x$ is indeed $\xi$--independent.  To
accomplish this we return to the integral equation (\ref{inteqm=mc})
and retain the general gauge dependence.  The equation can then
readily be converted to a differential equation:
\begin{eqnarray}
  \label{m=mde}
&&x \hat{G}''+ \left[ -1 - \frac{3}{2m^2 \pi^2 N} (1\!-\!\xi) x\hat{G}^2
  + \frac{3}{4\pi^2 N}
  (3\!-\!\xi) \frac{x}{x^2+m^2} \hat{G}^2 \right] \hat{G}' \nonumber\\ 
&&\qquad\qquad + \frac{1}{\pi^2 N} \left[ \frac{1}{m^2}
  - \frac{(5\!+\!\xi)}{4}\frac{1}{x^2+m^2} -
  \frac{2x^2}{\left(x^2+m^2\right)^2} \right] \hat{G}^3 = 0.
\end{eqnarray}
Considering the limit $x\rightarrow 0$ this reduces to a first order
differential equation (in which the term $x\hat{G}''$ has been dropped
as subleading in $x$) \[ \hat{G}' + \frac{(1\!+\!\xi)}{4\pi^2 m^2}
\hat{G}^3 =0\] which can be integrated easily to give for small $x$
\begin{equation}
  \hat{G}(x) \simeq \frac{1}{\left( c + \frac{(1\!+\!\xi)}{4\pi^2 N m^2}
  x \right)^{1/2}},
\end{equation}
demonstrating that the term in $\hat{G}''$ is indeed subleading and
also that all the $\xi$ dependence vanishes at $x=0$, so information
about the fixed point is $\xi$ independent as expected from the
on--shell nature of $\hat{G}(0)$.

We have shown in this section that in the presence of a mass for the
matter multiplet the full vertex is stabilized and driven to a
non--trivial infrared fixed point.  In the above analysis the mass
for the matter multiplet has been included by hand; however we will
show in section \ref{sec:mass} that a mass can be dynamically
generated self--consistently by coupling the vertex equation to the
corresponding Dyson--Schwinger equation for the matter propagator.  In
the next section we turn to a discussion of the inhomogeneous term $Z$
and why it can be safely omitted.

\section{The Inhomogeneous Term $Z$ And The Functions $A$ And $B$.}
\label{sec:z}

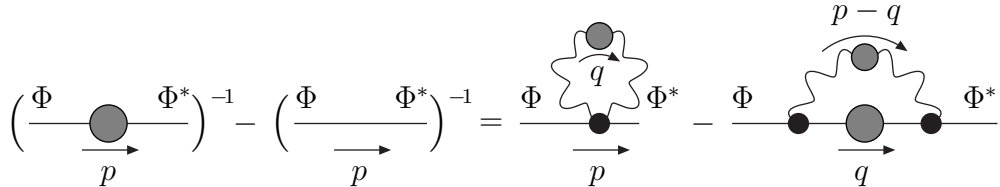
\begin{figure}[thb]
\begin{center}
\begin{picture}(370,80)(0,0)
\Text(0,30)[]{$\Big($}
\Line(5,30)(65,30)
\GCirc(35,30){7}{0.5}
\LongArrow(25,20)(45,20)
\Text(35,10)[]{$p$}
\Text(10,40)[]{$\Phi$}
\Text(60,40)[]{$\Phi^*$}
\Text(75,31)[]{$\Big)^{-\!1}$}
\Text(87,30)[]{$-$}
\Text(100,30)[]{$\Big($}
\Line(105,30)(155,30)
\Text(165,31)[]{$\Big)^{-\!1}$}
\LongArrow(120,20)(140,20)
\Text(130,10)[]{$p$}
\Text(110,40)[]{$\Phi$}
\Text(150,40)[]{$\Phi^*$}
\Text(180,30)[]{$=$}
\Line(190,30)(235,30)
\PhotonArc(220,48)(15,0,360){3.5}{7}
\GCirc(220,63){5}{0.5}
\LongArrow(210,20)(230,20)
\Vertex(220,30){4}
\LongArrowArcn(220,44)(12,130,50)
\Text(220,48)[]{$q$}
\Text(220,10)[]{$p$}
\Text(195,40)[]{$\Phi$}
\Text(245,40)[]{$\Phi^*$}
\Text(260,30)[]{$-$}
\Line(270,30)(370,30)
\GCirc(320,30){7}{0.5}
\PhotonArc(320,30)(25,0,180){4}{5.5}
\GCirc(320,55){5}{0.5}
\LongArrowArcn(320,38)(25,130,50)
\Text(322,72)[]{$p-q$}
\Vertex(295,30){4}
\Vertex(345,30){4}
\Text(275,40)[]{$\Phi$}
\Text(365,40)[]{$\Phi^*$}
\LongArrow(310,20)(330,20)
\Text(320,10)[]{$q$}
\end{picture}
\caption{\label{fig:dsm} Schematic form of the Dyson--Schwinger
  equation for the full matter propagator.  Solid lines represent
  matter superfield propagators, and wavy lines gauge superfield
  propagators; blobs indicate full non--perturbative quantities.}
\end{center}
\end{figure}

In this section we analyse the Dyson--Schwinger equations for the
functions $A$ and $B$ with which we have respectively dressed the
matter and connexion superfield propagators.  First we consider the
Dyson--Schwinger equation for the matter propagator, shown
schematically in figure \ref{fig:dsm}.

The difference between the graphs on the left of figure \ref{fig:dsm}
is easily computed to be
\begin{equation}
  \label{dslhs}
  -i (A(p)-1) \int d^2\theta \; \Phi(-p,\theta) D^2 (p) \Phi^*(p,\theta)
   -i \frac{M(p)}{A(p)} \int d^2\theta\; \Phi(-p,\theta) \Phi^*(p,\theta).
\end{equation}
We adopt the one--loop dressed approximation to the Dyson--Schwinger
equation in which both vertices in figure \ref{fig:dsm} are fully
dressed and all two--particle irreducible corrections are dropped
\cite{cornwall74}.  In doing so the resulting integral equation can be
cast in terms of $\hat{G}$ alone.

The graphs on the right hand side of figure \ref{fig:dsm} can be
manipulated to obtain functions multiplying the two superspace
structures
\begin{eqnarray}
  \label{suspstrs}
  &&\int d^2\theta\; \Phi(-p,\theta) \, D^2(p) \, \Phi^* (p,\theta)
  ,\nonumber\\
  &&\int d^2\theta\; \Phi(-p,\theta) \, \Phi^* (p,\theta);
\end{eqnarray}
comparison with equation (\ref{dslhs}) shows that the function
multiplying the first of these structures is to be identified with the
contribution to the wavefunction renormalization $A(p)$ and the
function which multiplies the second corresponds to the self energy
$M(p)/A(p)$, to which we will return in the next section.  The first
``seagull'' graph on the right hand side gives only an irrelevant
$p$--independent contribution to the wavefunction renormalization
while the last graph on the right hand side of the figure contributes
both to the wavefunction renormalization and the self energy.
Evaluating the superspace parts of the last graph leads to the
following integral equation for $A(p)$
\begin{eqnarray}
  A(p) = 1 + (1\!+\!\xi) A(p) \frac{ie^2}{2} \int \bard^3 k
  \frac{\hat{G}^2(k)}{k^2 + M^2(k)} \frac{k\cdot (k-p)}{(k-p)^4}.
\end{eqnarray}
In the gauge $\xi=-1$ which we adopt throughout this paper, this
reduces to unity, leaving a constant $A(p)$.  Note that this is in
line with the result of reference \cite{adrian99a}, where the
wavefunction renormalization was computed to be
\begin{eqnarray}
  A(p) = \left( \frac{p}{\alpha} \right)^{2(1\!+\!\xi)/N\pi^2}.
\end{eqnarray}

We turn now to the connexion superfield and the function $B(p)$.  In
standard large--$N$ treatments \cite{koopmans89,adrian99a} the
function $B$ includes the effects of massless matter loops resummed to
leading order in $1/N$, giving \[ B(p) \sim 1 + \alpha/p.\]  Therefore
in the deep infrared we find
\begin{equation}
  Z(x) = A^{-1} (x) \, B^{-1/2} (x) \sim x^{1/2}
\end{equation}
and the inhomogeneous term goes to zero and can be safely dropped
compared with a non--vanishing $\hat{G}(0)$ in a large--$N$ framework.

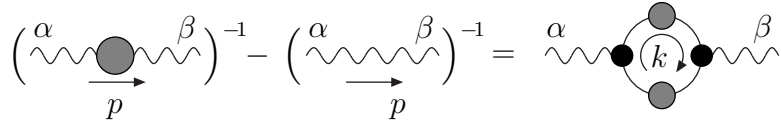
\begin{figure}[thb]
\begin{center}
\begin{picture}(280,60)
\Text(1,30)[]{$\Big($}
\Photon(6,30)(31,30){3}{2.5}
\GCirc(38,30){7}{0.5}
\Photon(45,30)(70,30){3}{2.5}
\Text(80,31)[]{$\Big)^{-\!1}$}
\LongArrow(28,20)(48,20)
\Text(38,10)[]{$p$}
\Text(11,40)[]{$\alpha$}
\Text(65,40)[]{$\beta$}
\Text(92,30)[]{$-$}
\Text(105,30)[]{$\Big($}
\Photon(110,30)(160,30){3}{4.5}
\Text(170,31)[]{$\Big)^{-\!1}$}
\LongArrow(125,20)(145,20)
\Text(145,10)[]{$p$}
\Text(115,40)[]{$\alpha$}
\Text(155,40)[]{$\beta$}
\Text(185,30)[]{$=$}
\Photon(200,30)(225,30){3}{2.5}
\CArc(244,30)(15,0,360)
\Text(244,30)[]{$k$}
\LongArrowArcn(244,30)(8,225,310)
\Photon(263,30)(288,30){3}{2.5}
\GCirc(229,30){4}{0}
\GCirc(259,30){4}{0}
\GCirc(244,45){5}{0.5}
\GCirc(244,15){5}{0.5}
\Text(205,40)[]{$\alpha$}
\Text(283,40)[]{$\beta$}
\end{picture}
\caption{Schematic form of the Dyson--Schwinger equation for the
  full connexion propagator.  Solid lines represent matter superfields
  and wavy lines represent connexion superfields; blobs indicate full
  non--perturbative quantities.}
\label{fig:dsc}
\end{center}
\end{figure}

For completeness it is instructive to consider the function $B$
outside this large--$N$ resummation, and consider the Dyson--Schwinger
equation shown schematically in figure \ref{fig:dsc}.  Again we adopt
the one--loop dressed approximation, which enables us to recast the
equation in terms of $\hat{G}$ alone.

The difference between the graphs on the left of figure \ref{fig:dsc}
is easily computed (in the gauge \(\xi=-1\)) to be
\begin{eqnarray}
  \label{dsclhs}
  \left( B(p) - 1 \right) \frac{1}{2} C^{\alpha\beta} p^2 \int d^2
  \theta\; \Gamma_\alpha (-p,\theta)\, \Gamma_\beta (p,\theta).
\end{eqnarray}
In computing the graph on the right of figure \ref{fig:dsc} the
superspace structure in equation (\ref{dsclhs}) naturally appears, and
the resulting integral equation for $B$ is (in which we have already performed
the simple angular integration)
\begin{eqnarray}
  p^3 \left(B(p) -1\right) = \frac{e^2 N}{16\pi^2} B(p) \int dk\;
  k\hat{G}^2(k) \, \ln \left( \frac{(k+p)^2 + M^2}{(k-p)^2+M^2} \right).
\end{eqnarray}
Converting to dimensionless variables, considering constant $M$ and
expanding the logarithm as usual we can convert to an equivalent
differential equation in which we have approximated $\hat{G}$ by its
constant value at $x=0$, noting from equation (\ref{vsoln}) that in
doing so we have only dropped terms of order ${\mathcal O}(x^3)$:
\begin{eqnarray}
  \frac{d}{dx} \left( x^2 B^{-1}(x) \right) =
  2x - \frac{\hat{G}^2(0)}{\pi^2} \frac{x^2}{m^2} +
  \frac{\hat{G}^2(0)}{\pi^2} \frac{x^2}{x^2+m^2}.
\end{eqnarray}
This can be integrated easily with result
\begin{eqnarray}
  B^{-1}(x) = 1 - \frac{1}{3} \frac{\hat{G}^2(0)}{\pi^2}
  \frac{x}{m^2} + \frac{\hat{G}^2(0)}{\pi^2} \frac{1}{x} -
  \frac{\hat{G}^2(0)}{\pi^2} \frac{m}{x^2} \arctan\left(\frac{x}{m}\right);
\end{eqnarray}
for finite solutions the constant of integration must vanish.  In the
limit of small $x$ there are cancellations to order ${\mathcal
  O}(x^3)$ and $B$ reduces to
\begin{eqnarray}
  B(x) \simeq \frac{1}{1 + {\mathcal O}(x^3)}.
\end{eqnarray}
From this analysis it is clear that even when the connexion propagator
is not resummed to leading order in $1/N$, the inhomogeneous term $Z$
reduces to a finite constant in the infrared and can still be dropped
with safety, justifying fully our assumption in section \ref{sec:dsv}.

\section{Mass.}
\label{sec:mass}

In this section we demonstrate that a mass for the matter multiplet
can be generated dynamically when we feed the solution for the full
vertex (\ref{vsoln}) into the Dyson--Schwinger equation for the matter
propagator.  The latter is shown schematically in figure
\ref{fig:dsm}, and now we evaluate the terms in the graph on the right
hand side which multiply the second of the superspace structures in
equation (\ref{suspstrs}).  Evaluating the superspace integrals leaves
the following integral equation for the mass function
\begin{eqnarray}
  M(p) = \frac{ie^2}{2} \int \bard^3 k \frac{\hat{G}^2(k)\,
  M(k)}{\left(k^2+M^2(k)\right) \, (k-p)^4} \left[ (1\!+\!\xi) k\cdot
  (k-p) + (1\!-\!\xi) (k-p)^2 \right].
\end{eqnarray}
Choosing again the gauge $\xi=-1$, we see that it removes the spurious
infrared divergences and simplifies the equation considerably.  After
performing the angular integration and expanding the resulting
logarithms, we obtain in dimensionless variables
\begin{eqnarray}
  m(x) = \frac{2}{\pi^2 N x^2} \int_0^x dy\; \frac{y^2 \, m(y)\,
  \hat{G}^2(y)}{y^2+m^2(y)} + \frac{2}{\pi^2 N} \int_x^\infty dy\; \frac{m(y)\,
  \hat{G}^2(y)}{y^2+m^2(y)}.
\end{eqnarray}
Differentiating with respect to $x$ this can be converted to an
equivalent differential equation,
\begin{eqnarray}
  \label{massde}
  x m'' + 3 m' + \frac{4}{\pi^2 N} \frac{m\,\hat{G}^2}{x^2+m^2} =0.
\end{eqnarray}
In the deep infrared $x\ll m$, and approximating $\hat{G}(x)$ by
$\hat{G}(0)$, noting again that the corrections to this approximation
are of order ${\mathcal O}(x^3)$ (see equation (\ref{vsoln}));
equation (\ref{massde}) then admits the following approximate
solution:
\begin{equation}
  \label{massdesoln}
  m(x) \simeq m(0) \left( 1 - a x -{\mathcal O}(x^2) \right) \qquad a>0
\end{equation}
which exhibits a constant dynamical mass $m(0)$ and has the correct
(decreasing) behaviour away from $x=0$; note that this solution is
similar to the small $x$ expansion of the mass found using the same
method for non--supersymmetric QED${}_3$ in reference \cite{mavro99}.
The solution above is to be compared with the case when the vertex is
chosen to satisfy the U(1) Ward identity for general momenta
\cite{adrian99a}, where the solution for the mass behaves as \[ m(x) =
m(0) e^{-x^2} \simeq m(0) \left( 1 - x^2 + {\mathcal O}(x^4)
\right).\]The semi--amputated vertex considered here only satisfies
the Ward identity at $x=0$ where the vertex is on--shell, and
therefore the small--$x$ dependence may differ from the more complete
solution, though the qualitative (decreasing) behaviour is retained.

\section{Conclusion.}
\label{sec:conc}

In this paper we have proposed that the dynamical generation of a mass
in \({\mathcal N}\!=\!1\) supersymmetric QED$_3$ is selected by the
dynamics over the massless alternative, for it stabilizes the running
gauge coupling against oscillations.  In particular the picture which
has emerged from our analysis is that in the absence of a mass for the
matter multiplet, the full vertex oscillates in the infrared, leading
to instability in the effective action.  When a finite dynamical mass
is added, the infrared physics is stabilized and there exists a
non--trivial infrared fixed point.  We have demonstrated that the
incorporation of our full vertex solution into the Dyson--Schwinger
equation for the matter propagator leads self--consistently to the
dynamical generation of a mass.

Supersymmetry has been kept manifest throughout by adopting a
superfield formalism which has the additional advantage of simplifying
considerably the system of equations and Ward identities which would
have to be solved in a component calculation.  By working in a general
gauge we have retained enough freedom to remove infrared divergences
which arise from the propagation of spurious degrees of freedom in the
connexion superfield \cite{clark77,clark+love}, and we have been able
to simplify the analysis significantly.  We have employed a large--$N$
framework consistently throughout this paper; this in combination with
the concept of a semi--amputated full vertex allowed us to decouple
the Dyson--Schwinger equations so that the functions $A$ and $B$ used
to dress the full propagators do not appear in the analysis of the
vertex.

There is the interesting possibility that the techniques employed in
this paper and in reference \cite{mavro99} could be extended to models
in higher dimensions and with non--Abelian gauge groups.

\section*{Acknowledgements.}

A.C.{--}S. would like to thank the CERN Theory Division for
hospitality, and gratefully acknowledges financial support from
P.P.A.R.C. (U.K.)  (studentship number 96314661), which made his visit
to CERN possible.  N.E.M. is partially supported by P.P.A.R.C. (U.K.)
under an Advanced Fellowship, and J.P. is funded by a Marie Curie
Fellowship (TMR--ERBFMBICT 972024).

\end{document}